\documentclass[pra,twocolumn,superscriptaddress,showpacs]{revtex4}
\usepackage{graphicx}
\def\op#1{\hat{#1}}
\def\vec#1{{\bf #1}}
\def\op#1{#1}
\def\ket#1{| #1 \rangle}

\def\ave#1{\langle #1 \rangle}
\def\norm#1{|| #1 ||}
\def\tr{{\rm Tr}}

\def\nat{Nature }
\newif\ifpdflatex\pdflatextrue
\makeatletter\@ifundefined{pdfoutput}{\pdflatexfalse}\makeatother
\def\myincludegraphics[#1]#2#3{%
\ifpdflatex \includegraphics[#1]{#2}
\else       \includegraphics[#1]{#3}
\fi}
\begin{document}
\bibliographystyle{prsty}
\title{Experimental Hamiltonian identification for controlled two-level systems}
\author{S. G. Schirmer} \email{sgs29@cam.ac.uk}
\affiliation{Department of Applied Maths and Theoretical Physics,
             University of Cambridge, Wilberforce Road, Cambridge, CB3 0WA, UK}
\affiliation{Department of Engineering, Division F, Control Group, 
             University of Cambridge, Trumpington Street, CB2 1PZ, UK}
\author{A. Kolli}
\affiliation{Department of Applied Maths and Theoretical Physics,
             University of Cambridge, Wilberforce Road, Cambridge, CB3 0WA, UK}
\author{D. K. L. Oi}
\affiliation{Department of Applied Maths and Theoretical Physics,
             University of Cambridge, Wilberforce Road, Cambridge, CB3 0WA, UK}
\date{\today}

\begin{abstract}
We present a strategy to empirically determine the internal and control 
Hamiltonians for an unknown two-level system (black box) subject to various 
(piecewise constant) control fields when direct readout by measurement is 
limited to a single, fixed observable.
\end{abstract}
\maketitle

Accurate determination of the dynamics of physical systems and their response
to external (control) fields is crucial for many applications, especially
quantum information processing (QIP). Quantum process tomography (QPT) is a
general procedure to identify the unitary (or completely positive) processes
acting on a system by experimentally determining the expectation values of a
complete set of observables~\cite{bib:QPT}.

However, QPT is only a partial solution in many cases.  One problem is the
assumption that one can experimentally determine the expectation values of 
a complete set of observables, or at least perform arbitrary single qubit
measurements for a register of $n$ qubits.  However, most QIP proposals
rely on a single readout process, i.e. measurement in one basis.  For example,
qubits encoded in internal electronic states of trapped ions~\cite{bib:ions} 
or neutral atoms~\cite{bib:atoms} are read out by quantum jump detection via
a cycling transition~\cite{bib:scatter}.  In solid-state systems based on
Cooper-pair boxes~\cite{bib:cooper}, Josephson junctions~\cite{bib:josephson},
or electrons in double-well potentials~\cite{bib:charge}, final readout is via
charge localization using single electron transistors~\cite{bib:SET} or similar
devices. Finally, solid state architectures based on electron or nuclear spin
qubits~\cite{bib:spin}, are expected to be limited to $\sigma_z$ measurements
via spin-charge transfer.

Generally, one assumes that a single projective measurement per qubit is
sufficient since arbitrary one-qubit measurements can be realized by performing
a local unitary transformation before measuring to achieve a change of basis.
However, implementing these basis changes requires accurate knowledge of the
dynamics of each individual qubit and its response to control fields in the
first place. Yet, in particular for solid-state
qubits~\cite{bib:cooper,bib:charge,bib:spin,bib:exitons,bib:josephson}, it is
difficult to \emph{predict} the actual dynamics of a qubit precisely based on
computer models and theory alone, since they may be sensitive to fabrication
variance and even vary from one qubit to the next as a result. We address this
problem of system identification with a single fixed measurement basis and
present a general strategy to identify the internal and control Hamiltonians
for a qubit subject a number of piecewise constant controls using a single,
fixed readout process. We can then bootstrap this process to implement QPT.

Although the technique may be adapted to dissipative systems, in this paper we
assume Hamiltonian evolution for the purpose of identifying the dynamics
relevant for the implementation of local unitary operations, i.e.  decoherence
times much greater than the gate operation times, which is crucial for systems
of interest in QIP.
The system evolution is thus governed by a unitary operator $\op{U}(t,t_0)$,
via $\op{\rho}(t)= \op{U}(t,t_0)\op{\rho}(t_0) \op{U}(t,t_0)^\dagger$, where
$\rho(t)$ is the system density operator, and $\op{U}(t,t_0)$ satisfies the
Schr\"odinger equation
\begin{equation} \label{eq:a}
  i\hbar \frac{d}{dt} \op{U}(t,t_0) = \op{H}[\vec{f}(t)] \op{U}(t,t_0),
\end{equation}
where $\op{H}$ is the Hamiltonian of the system.  The case of interest is when
$\op{H}$ depends on external fields $\{f_m\}$, which we can control.  Assuming
the fields are sufficiently weak and act independently, $\op{H}$ has the form
\begin{equation} \label{eq:b}
  \op{H}[\vec{f}(t)] = \op{H}_0 + \sum_{m=1}^M f_m(t) \op{H}_m,
\end{equation}
where $\op{H}_0$ is the free evolution Hamiltonian and $\op{H}_m$ for $m>0$
describes the interaction of the system with field $f_m$. Each $\op{H}_m$ can
in turn be expanded in terms of the Lie algebra generators; for a two-level 
system, the Pauli matrices $\op{\sigma}_j$ for
$j\in\{x,y,z\}$:
\begin{equation} \label{eq:c}
   2 \op{H}_m =   d_{m0} \op{I} +  d_{mx} \op{\sigma}_x 
              + d_{my} \op{\sigma}_y + d_{mz} \op{\sigma}_z.
\end{equation}
Thus we need to determine the real constants $d_{mx}$, 
$d_{my}$ and $d_{mz}$ for $m=0,1,\ldots,M$ (the $d_{m0}$ can be ignored since 
they result only in an unobservable global phase factor).

\begin{figure}
\myincludegraphics[width=3.3in]{figures/pdf/Fig1.pdf}{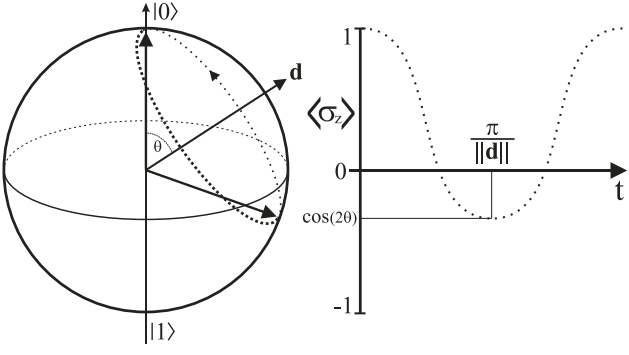}
\caption{Rotation about axis $\vec{d}$ and observable $z(\alpha_t)$.
  The expectation value of the measurement operator will oscillate
  sinusoidally, amplitude and frequency are directly related to the
  declination and length of the vector representing the applied Hamiltonian.}
\label{fig1}
\end{figure}

Geometrically, we can represent the state of the system by its Bloch vector
$\vec{s}(t)=(s_x,s_y,s_z)^T$ where $s_j=\tr[\op{\sigma}_j\op{\rho}(t)]$.  If
the system satisfies Eqs.(\ref{eq:a}-\ref{eq:c}), its Bloch vector evolves as
\begin{equation}
  \dot{\vec{s}}(t) = \left(R_0 + \sum_{m=0}^M f_m(t) R_m \right) \vec{s}(t)
\end{equation}
where $R_m$ are the real anti-symmetric rotation generators
\begin{equation}
  R_m = \left( \begin{array}{ccc}
               0 & d_{mz} & -d_{my} \\
               -d_{mz} & 0 & d_{mx} \\
               d_{my} & -d_{mx} & 0  
               \end{array} \right) 
\end{equation}
and $R=R_0+\sum_{m=1}^M f_m R_m$ generates a rotation about 
\begin{equation} \label{eq:d}
  \vec{d} = \vec{d}_0 + \sum_{m=1}^M f_{m} \vec{d}_m
\end{equation}
with $\vec{d}_m=(d_{mx},d_{my},d_{mz})^T$ for $m=0,1,\ldots,M$.  If $f_m(t)$
vary in time then so does $\vec{d}$.  If they are piecewise constant, having
fixed values $f_m$ for $t_0\le t\le t_1$, then $\vec{s}(t)$ for $t\in [t_0,t_1]$
rotates about the fixed axis $\vec{d}$, where $\norm{\vec{d}}$ is the rotation
frequency, and the unit vector $\hat{\vec{d}}=\frac{1}{\norm{\vec{d}}}
\vec{d}$ specifies the rotation axis. This allows us to give an explicit
formula for the trajectory of $\vec{s}(t)$ with $\vec{s}(0)=\vec{s}_0$:
\begin{eqnarray}
  \vec{s}(\alpha_t) 
  &=& \vec{s}_0 \cos\alpha_t 
      +\hat{\vec{d}}(\vec{s}_0 \cdot \hat{\vec{d}})(1-\cos\alpha_t) 
      +(\vec{s}_0 \times \hat{\vec{d}}) \sin\alpha_t \nonumber \\
  &=& [ \op{I} \cos\alpha_t + \op{A} (1-\cos\alpha_t)
                            + \op{B} \sin\alpha_t ] \vec{s}_0. \label{eq:s}
\end{eqnarray}
where the rotation angle is $\alpha_t = 
\alpha(t) =\norm{\vec{d}}(t-t_0)$ and 
\begin{eqnarray*}
 \op{A} &=& \frac{1}{2} \left(\begin{array}{ccc} 
  2\sin^2\theta\cos^2\phi & \sin^2\theta \sin(2\phi) & \sin(2\theta)\cos\phi \\
  \sin^2\theta\sin(2\phi) & 2\sin^2\theta\sin^2\phi  & \sin(2\theta)\sin\phi \\
  \sin(2\theta)\cos\phi   & \sin(2\theta)\sin\phi    & 2\cos^2\theta 
 \end{array} \right), \\
  \op{B} &=& \left(\begin{array}{ccc} 
           0                  & \cos\theta & -\sin\theta\sin\phi \\
           -\cos\theta        & 0          & \sin\theta\cos\phi \\
           \sin\theta\sin\phi & -\sin\theta\cos\phi & 0 
           \end{array} \right),
\end{eqnarray*}
and the unit vector $\hat{\vec{d}}$ is expressed in spherical coordinates,
$\hat{\vec{d}}=(\sin\theta\cos\phi,\sin\theta\sin\phi,\cos\theta)^T$.

Without loss of generality we shall assume that we can initialize the system
in the state $\vec{s}_0=\ket{0}=(0,0,1)^T$ with respect to the measurement
basis, and that we can experimentally determine the value of the $z$-component
of the Bloch vector $z=\ave{\op{\sigma}_z}$ by repeated measurements.
Inserting the expressions for $A$ and $B$ into Eq.(\ref{eq:s}), a rotation
about the axis $\vec{d}$ by the angle $\alpha_t=\norm{\vec{d}}(t-t_0)$
transforms the initial state $\vec{s}_0$ into $\vec{s}(\alpha_t)$ whose
$z$-component is
\begin{equation} \label{eq:z1}
  z(\alpha_t) = \cos(\alpha_t)\sin^2\theta + \cos^2\theta.
\end{equation}
$z(\alpha_t)$ is constant exactly if $\theta=0$.  Otherwise, it oscillates
and with minimum $z_{min} = \cos^2\theta-\sin^2\theta = \cos(2\theta)$ for
$\cos(\alpha_t)=-1$ or $\alpha_t=(2n+1)\pi$ (for some integer $n$).  Hence, 
if $\theta \neq 0$ then we can experimentally find $\norm{\vec{d}}$ and 
$\theta$ by determining the $t_{min}=\frac{\pi}{\norm{\vec{d}}}$ and $z_{min}$ 
(Fig.~\ref{fig1}).

\begin{figure}
\myincludegraphics[width=1.5in]{figures/pdf/Fig2.pdf}{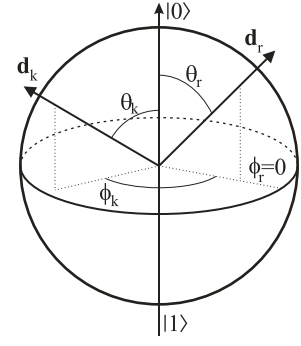}
\caption{Bloch sphere and arrangement of rotation axes. We can define the
  relative angular displacement of the vectors $\vec{d}_k$ representing
  applied Hamiltonians with respect to a reference vector $\vec{d}_r$
  ($\not=$z-axis).}
\label{fig2}
\end{figure}

For a single rotation it is sufficient to determine $\norm{\vec{d}}$ and
$\theta$ and set $\phi=0$.  For multiple rotations about different axes
$\vec{d}_k$, however, we must also determine the relative azimuthal angles
$\phi_k=\phi_k'-\phi_r'$ with respect to a fixed reference axis $\vec{d}_r$
(Fig.~\ref{fig2}).  The reference axis should have a vertical tilt angle
$\theta_r\in [\frac{\pi}{4}, \frac{3\pi}{4}]$ and we shall focus on this case
in this paper \footnote{If all axes have $\theta\not\in
  [\frac{\pi}{4},\frac{3\pi}{4}]$ then all we can do is map the initial state
  $\vec{s}_0$ to the state closest to the equatorial plane by a $\frac{\pi}{2}$
  rotation about the most horizontal axis and proceed in a similar manner as
  above.  However, the resulting expressions will be more complicated, and the
  accuracy of $\phi$ will diminish when $\vec{s}_1$ is close to $\vec{s}_0$,
  but such a system would be a poor candidate for a qubit.}.  Having determined
$\norm{\vec{d}}$ and $\theta$ of the axes, and chosen a
suitable reference axis $\vec{d}_r$, we can initialize the system in the state
$\vec{s}_1=(\cos\beta,\sin\beta,0)^T$ by rotating $\vec{s}_0$ about the axis
$\vec{d}_r$ by
$\alpha_r=\cos^{-1}\left(\frac{\cos(2\theta_r)+1}{\cos(2\theta_r)-1}\right)$,
where $\beta=\tan^{-1}(-\sqrt{-2\cos(2\theta_r)} \sec\theta_r)$.  Rotating
$\vec{s}_1$ by various angles $\alpha$ about the axis $\vec{d}_k$ then gives
$z(\alpha)= C(1-\cos\alpha)+D\sin\alpha$ with $C=\frac{1}{2}\sin(2\theta_k)
\cos(\phi_k-\beta)$ and $D=\sin\theta_k\sin(\phi_k-\beta)$, or equivalently
\begin{equation} \label{eq:z2}
  z(\alpha) = \gamma \sin(\alpha+\delta) - \gamma \sin\delta,
\end{equation}
where $C=\gamma\sin\delta$ and $D=\gamma\cos\delta$.  Hence, we can obtain
$\phi$ from experimental data for $z(\alpha)$ by finding $\delta$ and $\gamma$.

To determine the evolution of the system subject to the several external fields
we choose several field strengths $f_m^{(\ell)}$ ($\ell=1,\ldots,L$) for each 
control $f_m$ ($m=1,\ldots,M$) and compute the rotation axes 
\begin{equation}
  \vec{d}_{0+m}^{(\ell)} = \vec{d}_0 + f_m^{(\ell)} \vec{d}_m
\end{equation}
by finding $\theta_{0+m}^{(\ell)}$ and $\norm{\vec{d}_{0+m}^{(\ell)}}$ for all 
control settings, choosing a reference axis, and determining all the relative
azimuthal angles $\phi_{0+m}^{(\ell)}$ using the strategy outlined above.  Then
we plot the $x$, $y$ and $z$-components of $\vec{d}_{0+m}^{(\ell)}$ for $\ell=
1,2,\ldots,L$ versus the field strengths $f_m^{(\ell)}$ for each field, and fit
a straight line to each set of data points.  The vertical axis intercepts of the
lines then give the $x$, $y$, and $z$-component of $\vec{d}_0$, and their slopes 
determine the $x$, $y$, and $z$-component of $\vec{d}_m$.

To evaluate our strategy and find the best ways to extract the information from
noisy data we use computer simulations. We choose various sets of Hamiltonians,
use the proposed strategy to identify them and compare these results with the
actual values. Experiments are simulated by computing individual $\sigma_z$
measurements of $\vec{s}(t)$ according to Eq.(\ref{eq:s}), generating $N$
pseudo-random numbers $r_n\in[0,1]$, where $N$ is the number of times the
experiment is repeated, and taking the result of the $n^{\text{th}}$
measurement $M_n$ to be 1 if $r_n<(1+z)/2$ and 0 otherwise.  To account for
measurement errors we introduce a symmetric error probability $\eta\in [0,1]$
indicating the frequency of erroneous measurement results, i.e., registering 1
when we should have measured 0, and vice versa.  This is achieved by choosing
another $N$ random numbers $e_n \in [0,1]$, and changing $M_n$ to $1-M_n$
whenever $e_n<\eta$.  Finally, we set $\ave{\sigma_z}=-1+\frac{2}{N}
\sum_{n=1}^N M_n$.

We illustrate our Hamiltonian identification strategy in detail with a specific
test system:
\begin{equation}
  \begin{array}{l}
  2\op{H}_0 = 0.2\sigma_x + 0.1\sigma_z, \;
  2\op{H}_1 = \sigma_x + 0.9 \sigma_y + 0.1 \sigma_z, \\
  2\op{H}_2 = 0.2\sigma_x + 0.9\sigma_z.
  \end{array}
\label{eq:testsystem}
\end{equation}
This Hamiltonian may arise, for instance, for a charge qubit \cite{bib:charge}
consisting of two quantum dots sharing a single electron, and two voltage gates 
$B$ and $S$, intended to enable us to change the potential barrier between them
and induce asymmetries in the double well potential. The measurement basis states
$\ket{0}$ and $\ket{1}$ are the localized ground states of the charge in either 
of the two dots, and projective measurements onto them can be performed via a 
single electron transistor placed next to one of the dots. $\op{H}_0$ indicates
some tunneling between the dots when no control voltages are applied to the 
gates and a slight potential asymmetry, which might be due to imperfect placement
of the donor impurities.  $\op{H}_1$ induces tunneling between the dots as 
desired but also induces a slight potential asymmetry, which might occur if the
electrode was placed slightly off-center. $\op{H}_2$ produces the desired energy
level shift but also increases tunneling slightly.

\begin{figure}
\myincludegraphics[width=3.3in]{figures/pdf/Fig3.pdf}{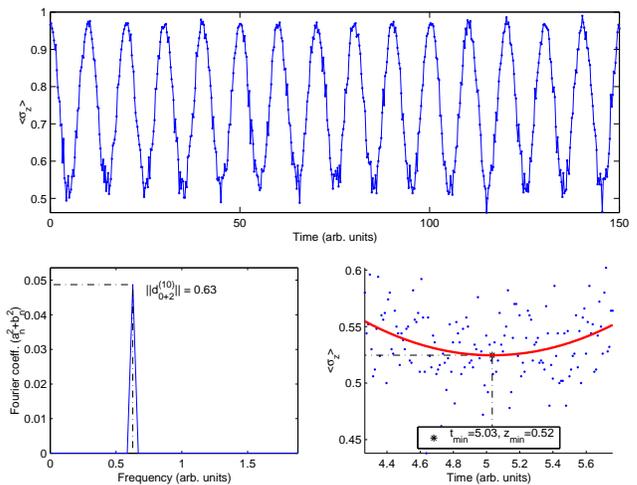}
\caption{Identification of $\norm{\vec{d}}$ and $\theta$ for $\vec{d}=\vec{d}_0
  +f_2^{(10)}\vec{d}_2$. Shown is simulated data for $\vec{d}_{0+2}^{(10)}$
  for $f_2^{(10)}=0.5$.  The top graph is used to estimate the rotation
  frequency using the DFT, the bottom left plot shows the Fourier transform,
  and the bottom right plot the magnification of the region of interest as
  well as the parabola that provides the best least-squares fit to the data.}
\label{fig3}
\end{figure}

The first step towards identifying the Hamiltonians $\op{H}_0$, $\op{H}_1$ and
$\op{H}_2$ involves finding the rotation frequencies $\norm{\vec{d}_0}$ as
well as $\norm{\vec{d}_{0+1}^{(\ell)}}$, $\norm{\vec{d}_{0+2}^{(\ell)}}$ and
the angles $\theta_0$, $\theta_{0+1}^{(\ell)}$, $\theta_{0+2}^{(\ell)}$ for
several gate voltages $f_1^{(\ell)}$ and $f_2^{(\ell)}$.  We do this by
applying each gate voltage for various periods of time $t_j$ and finding
$z(t_j)=\ave{\sigma_z(t_j)}$ (Fig.~\ref{fig3}).  An estimate of the rotation
frequency $\omega$ is obtained by taking the discrete Fourier transform (DFT)
of $z(t_j)-\frac{1}{J}\sum_{j=1}^J z(t_j)$ and identifying its maximum. We
then estimate the time $t_*=\frac{\pi}{\omega}$ when $z(\alpha_t)$ is expected
to assume its first minimum, acquire (circa 150) additional data points for
$t\in [0.85t_*,1.15t_*]$ and determine the minimum $z_{min}$ by fitting a
\emph{parabola} to these data points; $\theta=\frac{1}{2}\cos^{-1}(z_{min})$.
This proved to be considerably more accurate for noisy data than fitting a
function of the form Eq.(\ref{eq:z1}) to the data points and reading off the
frequency and minimum from this curve directly.

\begin{figure}
\myincludegraphics[width=3.3in]{figures/pdf/Fig4.pdf}{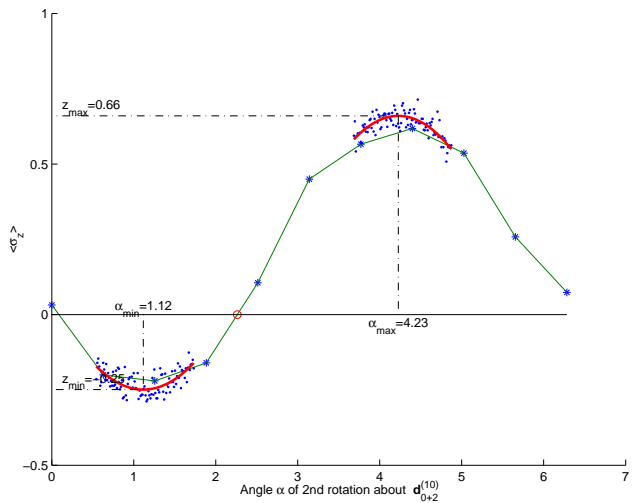}
\caption{Identification of $\phi$ for $\vec{d}=\vec{d}_0+f_2^{(10)}\vec{d}_2$.
  The points $*$, are a coarse sampling of the curve from which the first axis
  crossing can be estimated (red circle). This then gives estimates of the
  turning points $\alpha_{min}^{est}$ and $\alpha_{max}^{est}$ which can be
  refined by re-sampling (dots) in their vicinity and fitting a parabola (red
  curves). From this we determine the vertices $(\alpha_{min},z_{min})$ and
  $(\alpha_{max},z_{max})$.}
\label{fig4}
\end{figure}

Next we determine the angles $\phi_{0+m}^{(\ell)}$ relative to the reference
axis, chosen to be $\vec{d}_0$ here. In all of the following experiments the
system is initialized in state $\vec{s}_1=(\cos\beta,\sin\beta,0)^T$ by
applying a suitable rotation about the reference axis. In our case the state
$\vec{s}_0$ is rotated to $\vec{s}_1$ by letting it evolve freely for $\approx
8.1379$ time units yielding $\beta \approx -1.0489$.  We then apply each field
$f_m^{(\ell)}$ for various times to achieve rotations by various angles
$\alpha$ (Fig.~\ref{fig4}).
This yields the parameters $\gamma=\frac{1}{2}(z_{max}-z_{min})$ and
$\delta=\pi-\frac{1}{2}(\alpha_{min}+ \alpha_{max})$ in Eq.(\ref{eq:z2}) from
which we can obtain $\phi$, e.g., by setting $D=\gamma\cos\delta$ and
$\phi=-\beta-\sin^{-1}(D/\sin(\theta))$ where $\theta$ is the vertical tilt
angle of the rotation axis determined in the previous step.

\begin{figure}
\myincludegraphics[width=3.3in]{figures/pdf/Fig5.pdf}{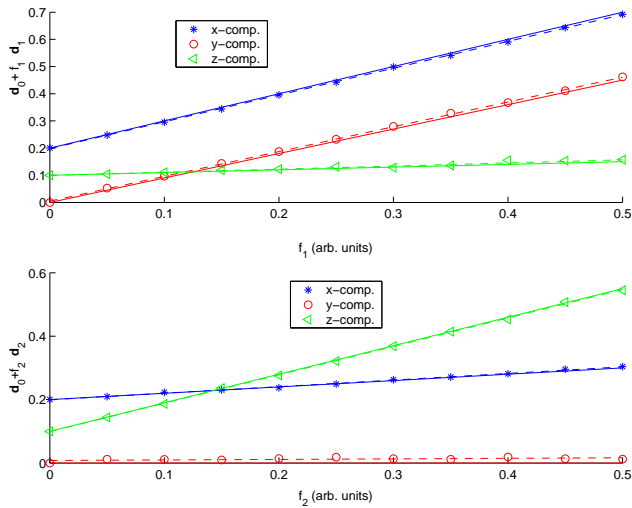}
\caption{Identification of the components of $\vec{d}_0$ and $\vec{d}_m$.
  Typical simulated experimental run for the system defined by
  Eq.(\ref{eq:testsystem}) and the estimated components of $\vec{d}$ for
  different field settings. Shown are the best straight line fit (dashed
  lines) and the actual values (solid line). Note the excellent agreement
  between the actual values and the best fit obtained.  A typical run of the
  program yielded: $\vec{d}_0=(0.1987,0.0064,0.0992)^T$
  $\vec{d}_1=(0.9859,0.9122,0.1149)^T$, $\vec{d}_2=(0.2081,0.0170,0.8957)^T$,
  i.e., the Hilbert-Schmidt norm errors
  $\norm{\vec{d}_m^{est}-\vec{d}_m^{act}}$ were less than 3\% ($0.0066$,
  $0.0238$ and $0.0193$ for $m=0,1,2$ respectively).}
\label{fig5}
\end{figure}

Finally, having determined all the relevant parameters of the rotation axes,
we convert the data to Cartesian coordinates, and plot the $x$, $y$ and $z$
components of $\vec{d}_{0+m}^{(\ell)}$ for each field $f_m$ as a function of
$f_m^{(\ell)}$ (Fig.~\ref{fig5}). From the straight line fitting, we can then
estimate the Hamiltonians.

In conclusion, we have presented a stepwise procedure to determine the dynamical
response of a Hamiltonian 2-level quantum system to control fields when we only
have access to measurement in a single fixed basis.  This overcomes a weakness 
in the assumptions of QPT and thus enables us to bootstrap the procedure to 
estimate parameters for systems consisting of arrays of qubits.  Numerical results
indicate that the procedure is fairly robust for Hamiltonians of physical interest.
Extensions of the procedure higher-dimensional or dissipative systems may be 
possible.

SGS and DKLO acknowledge funding from the Cambridge-MIT Institute, EU grants
RESQ (IST-2001-37559), TOPQIP (IST-2001-39215) and Fujitsu.


\begin{thebibliography}{99}
\bibitem{bib:QPT}
I.~L.~ Chuang, M.~A.~ Nielsen, \jmo {\bf 44}, 2455 (1997);
J.~F.~Poyatos, J.~I.~Cirac, P.~Zoller, \prl {\bf 78}, 390 (1997)

\bibitem{bib:ions}
J.~I.~Cirac, P.~Zoller, \prl {\bf 74}, 4091 (1995) \& \nat {\bf 404}, 579 (2000);
D.~Kielpinski, D.~C.~Monroe, D.~J.~Wineland, \nat {\bf 417}, 709 (2000)

\bibitem{bib:atoms}
I.~H.~Deutsch, G.~K.~Brennen, P.~S.~Jessen, Prog.\ of Phys.~{\bf 48}, 925 (2000);
H.~J.~Briegel \textit{et al.~}, 
\jmo {\bf 47}, 415 (2000); C.~Monroe, \nat {\bf 416}, 238 (2002).~

\bibitem{bib:scatter}
W.~Nagourney, J.~Sandberg, H.~Dehmelt, \prl {\bf 56}, 2797 (1986); 
Th.~Sauter \textit{et al.}, 
\prl {\bf 57}, 1696 (1986);
J.~C.~Bergquist \textit{et al.}, 
\prl {\bf 57}, 1699 (1986)

\bibitem{bib:cooper} 
Y.~Maklin, G.~Sch\"on, A.~Shnirman, \nat {\bf 398}, 305 (1999); 
Y.~Nakamura, Yu.~A.~Pashkin, J.~S.~Tsai, {\it ibid.~}, 786.~

\bibitem{bib:josephson}
Y.~Maklin, G.~Schon and S.~Shnirman, \rmp {\bf 73}, 357 (2001);
Y.~Nakamura, Y.~A.~Pashkin and J.~S.~Tsai, \prl {\bf 87}, 246601 (2001)

\bibitem{bib:charge}
L.~C.~L.~Hollenberg \textit{et al.}, cond-mat/0306235;
A.~S.~Dzurak \textit{et al.}, cond-mat/0306285 (2003)

\bibitem{bib:SET}
M.~H.~Devoret, R.~J.~Schoelkopf, \nat {\bf 406}, 1039 (2000); 
A.~A.~Clerk \textit{et al.}, 
\prl {\bf 89}, 176804 (2002);
T.~M.~Buehler \textit{et al.}, cond-mat/0304384 (2003);
J.~Kinnunen, P.~Torma, J.~P.~Pekola, cond-mat/0211154 (2002).~

\bibitem{bib:spin}
D.~Loss and D.~P.~DiVincenzo, \pra {\bf 57}, 120 (1998);
B.~E.~Kane, \nat {\bf 393}, 133 (1998)

\bibitem{bib:exitons}
A.~Ekert, R.~Josza, \rmp {\bf 68}, 733 (1996)



\end{thebibliography}
\end{document}